\documentclass[10pt,journal,compsoc]{IEEEtran}

\usepackage{amsmath}
\usepackage{amssymb}
\usepackage{algorithmic}
\usepackage{cite}	
\usepackage{graphicx}
\usepackage{amsthm}
\usepackage{mathtools}
\usepackage{algorithm}
\usepackage{algorithmic}
\usepackage{xcolor}
\usepackage{siunitx}
\ifCLASSOPTIONcompsoc
\usepackage[caption=false, font=normalsize, labelfont=sf, textfont=sf]{subfig}
\else
\usepackage[caption=false, font=footnotesize]{subfig}
\fi
\newtheorem{thm}{Lemma}
\newtheorem{prop}{Proposition}

\begin{document}
\title{Hybrid Coded-Uncoded Caching in Multi-Access Networks with Non-uniform Demands }

\author{\IEEEauthorblockN{Abdollah Ghaffari Sheshjavani\IEEEauthorrefmark{1}, 
Ahmad Khonsari\IEEEauthorrefmark{1}\IEEEauthorrefmark{2},
Masoumeh Moradian\IEEEauthorrefmark{2},	
Seyed Pooya Shariatpanahi\IEEEauthorrefmark{1}, 
and Seyedeh Bahereh Hassanpour\IEEEauthorrefmark{1},  \\ 
}
\thanks{\IEEEauthorblockA{\IEEEauthorrefmark{1}School of Electrical and Computer Engineering, College of Engineering, University of Tehran, Iran}}
\thanks{\IEEEauthorblockA{\IEEEauthorrefmark{2}School of Computer Science, Institute for Research in Fundamental Sciences (IPM), Iran}}
\thanks{Emails: \{abdollah.ghaffari, a\_khonsari, p.shariatpanahi\}@ut.ac.ir, mmoradian@ipm.ir}
}

\maketitle

\begin{abstract}
\fontdimen2\font=0.54ex
To address the massive growth of data traffic over cellular networks, increasing spatial reuse of the frequency spectrum by the deployment of small base stations (SBSs) has been considered. For rapid deployment of SBSs in the networks, caching popular content along with new coded caching schemes are proposed.
To maximize the cellular network's capacity, densifying it with small base stations is inevitable. In ultra-dense cellular networks, coverage of SBSs may overlap. To this aim, the multi-access caching system, where users potentially can access multiple cache nodes simultaneously, has attracted more attention in recent years. 
Most previous works on multi-access coded caching, only consider specific conditions such as cyclic wrap-around network topologies. In this paper, we investigate caching in ultra-dense cellular networks, where different users can access different numbers of caches under non-uniform content popularity distribution, and propose Multi-Access Hybrid coded-uncoded Caching ($MAHC$). We formulate the optimization problem of the proposed scheme for general network topologies and evaluate it for 2-SBS network scenarios. The numerical and simulation results show that the proposed $MAHC$ scheme outperforms optimal conventional uncoded and previous multi-access coded caching ($MACC$) schemes.
\end{abstract}


\section{Introduction}
Global Internet traffic has grown continuously at a compound annual growth rate (CAGR) of over 30\% in recent years. The growth of mobile network traffic has been faster than that, where the share of mobile network traffic of the total Internet traffic has increased from 15\% in 2017 to about 30\% in 2022. Additionally, demand for receiving content through wireless media is constantly increasing, as expected that mobile network traffic increase from 115 exabytes per month in 2022 to 452 exabytes per month in 2028 \cite{2022ericsson}.
A promising solution to mitigate this massive mobile data traffic is increasing spatial reuse of the frequency spectrum by shrinking the network cell sizes through the deployment of small base stations (SBSs) \cite{chandrasekhar2008femtocell}. Nonetheless, the high cost and time-consuming creation of wired links and the bottleneck of wireless links are the main obstacles to the rapid deployment of SBSs in networks. Many works in the literature address this problem and propose to use of wireless media for rapid deployment SBSs, and caching popular contents at them to relieve the need for high-speed backhaul links \cite{shanmugam2013femtocaching, chen2017probabilistic}. 

There are three basic caching schemes in the literature: conventional (uncoded) caching, coded caching, and hybrid coded-uncoded caching. In conventional caching instead of fetching the popular contents from the original server every time a user requests them, it is desirable to store them in the proximity of end-users and serve locally. Caching more popular contents at local caches, i.e., the popularity principle, leads to the so-called local caching gain, which is proportional to the local memory size. In \cite{maddah2014fundamental}, a more recent scheme called coded caching is proposed that significantly improves performance over conventional caching by using a carefully designed cache placement of uncoded contents and a coded delivery, which leverages the multi-casting nature of the shared (such as wireless) medium. In this scheme, several requested contents are coded in sub-files and sent over the shared medium. Then, each user uses its' cached contents to decode the desired content from the coded sub-files. In coded caching, it is desirable to cache diverse parts of the library among different caches to increase multi-casting opportunities, i.e., the diversity principle. This, in addition to the local caching gain, results in the global caching gain, which is proportional to the aggregate memory of all the caches. These two diversity and popularity principles are in tension when the content popularity distribution is non-uniform. Finally, the hybrid coded-uncoded caching scheme is proposed in \cite{9120820, ghaffarisheshjavani2021content},which significantly outperforms the pure coded and conventional pure uncoded schemes, mainly when each cache receives multiple requests. In the original works of hybrid coded-uncoded caching \cite{9120820, ghaffarisheshjavani2021content,9839153}, the cache of SBSs is divided into two parts; coded and uncoded and optimal strategy, i.e., the optimal cache partitioning and contents selection, is found by trading-off between popularity and diversity.

Former works on caching in telecommunication networks assumed that each user had access to only one cache-equipped station. Recently, to satisfy the massive growth of mobile data traffic, densifying the cellular networks with small base stations is considered. In ultra-dense cellular networks, coverage of SBSs may overlap, and users can potentially access multiple SBSs simultaneously. Therefore, the multi-access caching system has attracted the attention of researchers on the subject of conventional and coded caching schemes \cite{8374848,9638333, multilevel2017}.

When users access more than one cache simultaneously, the tension between popularity and diversity is affected. Generally, in dense cache networks, where users have access to more cache nodes, storing more diverse contents in the cache nodes improves the caching performance. Even for conventional pure uncoded caching, in dense networks, caching diverse contents in different caches can improve the caching performance. Otherwise, if users have access to fewer cache nodes, it is better to cache more popular contents in the caches. Therefore in multi-access caching problems, network topology should also be considered to find the optimal caching strategy, which makes the problem more complicated. 

Most previous works on multi-access coded caching ($MACC$) consider specific network topologies, such as having exactly $r>1$ number of connections to caches for each user with cyclic wrap-around connectivity \cite{multilevel2017,9036921,9774404, 9511442,8989128,9551929,10068305,9771663,9834460}. These assumptions are not realistic, especially in cellular networks, where network topology is non-stationary due to the mobility of users. In addition, previous works in $MACC$ do not consider non-uniform content popularity distribution.
Therefore, in this paper, we consider multi-access caching under non-uniform content popularity distribution in general network topology, where users can access different numbers of caches, and propose and formulate the Multi-Access Hybrid Caching ($MAHC$) by modifying the baseline hybrid coded-uncoded caching scheme proposed in \cite{9120820, ghaffarisheshjavani2021content}. We analyze and evaluate multi-access caching in the case of 2-SBS scenarios. The numerical and simulation results show that the proposed $MAHC$ outperforms optimal conventional uncoded and $MACC$ schemes.

The rest of the paper is organized as follows. In Section~\ref{sec2} an overview of the coded and hybrid coded-uncoded caching is provided, and the related works are reviewed. We introduce the system model in Section~\ref{sec3}. The proposed caching schemes, problem formulation, and performance analysis are described in Section~\ref{PS}. This is followed by numerical analysis and simulation results in Section~\ref{sec6}. Finally, Section~\ref{cocln} concludes the paper.
\section{ Background  and Related Works}
\label{sec2}
In this section, we first summarize the baseline coded caching scheme reported in \cite{maddah2014fundamental} and then review the hybrid coded-uncoded caching\cite{ghaffarisheshjavani2021content,9120820}. Finally, we survey the related works on multi-access scenarios.

\subsection{Background on coded and hybrid coded-uncoded caching}
\label{hybrid_Back}
In the original coded caching scheme\cite{maddah2014fundamental} one server is connected through a shared, error-free link to $K$ number of users. The server have access to the library of $N$ contents each of size $F$ bits. Each user is equipped with a cache memory of size $MF$ bits. The system operates in two phases: a $placement$ phase and a $delivery$ phase. 
In the placement phase, each content splits into ${K}\choose{T}$ non-overlapping equal-sized sub-files, where $T=K \!\times\! M / N$ and the size of each sub-file is equal to $F/{{K}\choose{T}}$. The sub-files are distributed at caches such that each cache stores $M/N$ of each content. Moreover, each sub-file has $T$ copies in $T$ different caches. In the delivery phase, each cache receives a request for a single content. Then, the server XORs the required sub-files according to a specific coding strategy and multi-casts coded messages to the corresponding groups of $T+1$ caches. The achievable rate of the coding strategy for serving all contents at the shared link is proven to be \cite{maddah2014fundamental}:
\begin{equation}
	R = K \left(1-\frac{M}{N}\right) \min \left\{\frac{1}{1+K\times M/N},\frac{N}{K}\right\}.
	\label{eq2}
	\vspace{-0.3em}
\end{equation}
 This idea later generalized to hierarchical coded caching \cite{karamchandani2016hierarchical}, multi-server coded caching \cite{shariatpanahi2016multi}, online coded caching\cite{7055939}, decentralized coded caching \cite{7999228}, device-to-device (D2D) coded caching  \cite{7342961}, coded caching with asynchronous user requests \cite{8374865}, and coded caching with multiple file requests \cite{8594642}. Although the original coded caching scheme \cite{maddah2014fundamental} performs well under uniform content popularity, this scheme is inefficient in non-uniform content popularity. To handle non-uniform content popularity, grouping contents based on their popularity and treating each group as a single coded caching problem has been proposed in the literature\cite{niesen2017coded, li2017traffic, ji2017order, zhang2018coded, multilevel2017, 8863425}.
 
   More recently, in \cite{ghaffarisheshjavani2021content,9120820}, we show that when each cache receives multiple requests in the delivery phase, partitioning contents into (at most) three uncoded-cached, coded-cached, and non-cached groups and using hybrid conventional uncoded and coded caching outperforms the baseline schemes of pure uncoded and pure coded caching.

In the placement phase of the baseline hybrid coded-uncoded caching scheme, each cache memory is divided into two parts: $M_1 \!\times \!F$ and $(M \!-\! M_1) \!\times F$ bits. Accordingly, $N_1$ most popular contents have been chosen and among them, $M_1$ most popular ones are cached at all caches entirely. The $N_1 \!-\! M_1$ contents are cached using the original coded caching scheme proposed in \cite{maddah2014fundamental}. The remaining $N-N1$ less popular contents are not cached at all. In the delivery phase, each cache $c$  receives $Z_c$ requests, where $c \in \{ 1,2, \dots, K\}$. The requests from the $M_1$ most popular contents locally are served by caches, and the server is responsible for the requests for coded-cached and un-cached contents. In order to perform the coded caching scheme, the server maintains the requests for $N_1-M_1$ coded contents of each cache separately in $K$ distinct queues. The server transmits the $i$th rows of all coded queues in step $i$ by the coded scheme. The number of queues involved in the coding process at step $i$ could be less than $K$ since some of the queues may not have any requests at step $i$. Moreover, the number of steps is at most $\max_{c\in\{1,\dots, K\}}\{Z_c\}$. Finally, after sending all coded requests, the server handles requests for un-cached contents. The achievable rate of the hybrid caching strategy for serving all contents at the shared link is proven to be $r_1+r_2$ where:
\begin{align}
		\label{lem2-r1}
		r_{1}=
		\begin{cases}
			\scalebox{.95}{$F \times \sum_{i=1}^{Z_{max}} \frac{{{K}\choose {T+1}}-\sum_{k=0}^{K}Pr\{Q_i=k\}{{K-k}\choose {T+1}}}{{{K} \choose{T}}}, \quad if\ N_1>M,$} \\
			0,\quad otherwise
		\end{cases}   
	\end{align}
	\begin{align}   
	r_{2}=F \times \sum_{n={N_1}+1}^{N} 1-(1-p_{n})^{\sum_{c=1}^{K}Z_c}.
	\label{proposition2-r2}
	\vspace{-0.4em}
	\end{align}
	In the above equations,  $r_1$ is the expected traffic rate of coded content requests, and $r_2$ is the expected traffic load of the uncoded content requests. In addition, $Z_{max}= \max_{c} Z_c$, where $c \in \{ 1,2, \dots, K\}$, and $Pr\{Q_{i}=k\}$ denotes the probability that exactly $k$ queues contain requests at step $i$. 
\subsection{Related Works}
\label{RW}
  Coded caching has been investigated in different network scenarios such as shared medium networks and Device-to-Device (D2D) communication networks. $MACC$ is one the most interesting scenarios which is getting popular recently. In this part, we review the previous works which have been done in $MACC$ scenarios.

 	Many previous works in $MACC$ consider $K$ users and $K$ caches where each user is connected to exactly $r>1$ number of caches according to a specific network topology called ``cyclic wrap-around’’, in which user $k$ has access to caches $k, k+1, k+2, \dots, k+r-1$ for $r \in \{1,2,\dots,K\}$. $r$ is denoted as the cache degree which defines the number of caches a user is connected to \cite{multilevel2017,9036921,9774404, 9511442,8989128,9551929,10068305,9771663,9834460}. 
	
	Authors in \cite{multilevel2017} considered $MACC$ with cyclic wrap-around network topology in a distributed setting 
	where the cache degree $r$ divides $K$. They leveled users into $U$ different levels based on their popularities. Moreover, they investigate the rate memory trade-off for multi-level popularity and access and derived the achievable rate and information-theoretic lower bound. Following that, authors in \cite{9036921, 9774404} derived a lower bound on the optimal rate-memory trade-off, which is tighter than the lower bounds in \cite{multilevel2017}.The $MACC$ scheme proposed in \cite{multilevel2017} suffers from the case that $r$ does not divide $K$. Authors in \cite{9511442} considered this problem and proposed a new $MACC$ scheme by applying a transformation approach to the original shared-link coded caching scheme \cite{maddah2014fundamental}. The result shows that this approach achieves the same load as the scheme of \cite{multilevel2017} but for any system parameters.

	Authors in \cite{8989128} also considered $MACC$ with cyclic wrap-around network topology and proposed a caching and coded delivery scheme which achieved a coding gain higher than $KM/N +1$ for two opposing memory regimes. In \cite{9551929,10068305}, the authors propose a content placement strategy with linear sub-packetization from the perspective of the Placement Delivery Array (PDA). In \cite{9551929}, the authors construct a new class of PDA called t-cyclic g-regular PDA and studied it in the case of $T=KM/N$  where $K$ divides $T$, denoted by $T|K$, and $( K-(T \times L) +T)|K$.	In \cite{10068305}, the authors propose a content placement strategy called consecutive cyclic placement and derive the optimal coded caching gain. The proposed scheme has a better coded caching gain compared to some previous schemes with linear sub-packetization such as \cite{9551929}.
	
	Authors in \cite{9771663,9834460} considered the privacy of demands in $MACC$ with cyclic wrap-around network topology, where each user can only access its required file and cannot obtain any information about the demands of other users. Authors in \cite{9834460} formulated a multi-access caching system with privacy demands and proposed an approach to guarantee privacy for existing non-private $MACC$ schemes by storing some private keys at the cache nodes.
	
	Authors in \cite{9834838} considered a two-dimension cache and user placement where there are $K_1$ and $K_2$ caches and every user has access to $r^2$ neighboring caches. They show that for a specific ultra-dense multi-access coded caching, instead of caching diverse pieces of a content, grouping caches and caching diverse contents in the caches which serve the same users, slightly reduces the load on the shared medium. 
	
	Works so far mentioned above considered a specific network topology with cyclic wrap-around. Some other $MACC$ works consider scenarios with a larger number of users compared with the number of caches. These works assume $K$ number of users and $B$ number of caches, where also each user is  connected to exactly $r>1$ number of caches but in a generalized network topology \cite{katyal2021multi, 9611394 ,9839411}.

Authors in \cite{katyal2021multi} considered connection between coding and topology and developed a $MACC$ scheme from a specific type of resolvable design called cross-resolvable designs which is a class of combinatorial design. In order to provide a comparison with other approaches proposed in the literature, authors apply a per-user-rate approach. Results denote the outperformance of the proposed approach in \cite{katyal2021multi} compared to the \cite{8989128}. Similar to the last settings, authors in \cite{9611394} provide a general model of the original coded caching scheme \cite{maddah2014fundamental} for $MACC$ and evaluate the performance of the proposed approach with \cite{multilevel2017} and \cite{8989128}. Finally, authors in \cite{9839411} proposed a $MACC$ scheme for a general scenario of the entire ensemble of all possible network connectivity/topologies through an information-theoretic approach. 
\section{System Model}
\label{sec3}
\begin{figure}[!t]
	\centering
	\includegraphics[width=3.2in]{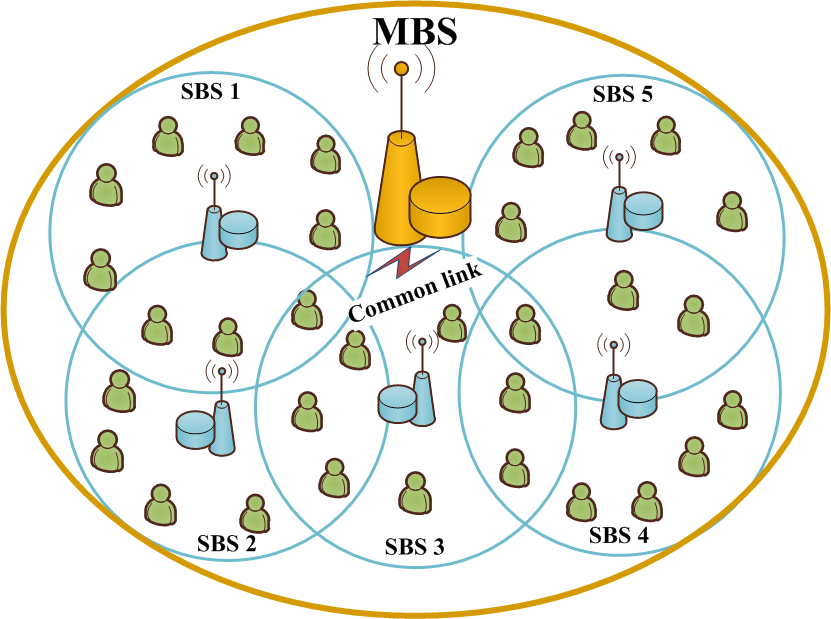}
	\caption{Illustrative system model of caching in dense cellular networks.}
	\label{SMFigure}
	\vspace{-0.5em}
\end{figure}

We consider a cellular network consisting of one MBS, which is connected through a shared error-free link to $K$ SBSs, as depicted in \figurename{~\ref{SMFigure}}. The content library has $N=\{W_1, W_2, ... , W_N\}$ distinct contents that are all accessible by the MBS.  Without loss of generality, we assume that all contents have the same size equal to $F$ bits. Each SBS $c$ has a cache memory of size $M \times F$ bits for some integer number of $M \in [0, N]$. There are $Z$ users in which each user can connect and receive data from the MBS and one or more SBSs. SBSs and users spread in the area of MBS arbitrarily. Although the network topology (i.e., connectivity relations between SBSs and users) is arbitrary, a proper estimation of SBSs' regions and their intersections  (and the number of users in each area) is available. 
Similar to previous works, our system operates in two phases: the content placement phase and the content delivery phase. The placement phase is carried out during off-peak times, where the caching strategy determines some functions of all contents $s_c=f_c(W_1, ..., W_N)$, $c \in \{ 1, 2, \dots, K\}$ which must be cached at each SBS based on the system parameters such as cache memory constraint, network topology (connectivity), and content popularity. In this phase, the caches are filled with corresponding contents from the library.  
We assume that the content popularity distribution is arbitrary and  does not change during the delivery phase.  $p_{n}$ denotes the probability of requesting the content $W_n$, where $n \in \{1,2, \dots, N\}$. 

In the delivery phase, users reveal their requests and only the MBS accesses the whole library. We assume each user has one request at each delivery phase, resulting in a total of $Z$ requests. It is worth noting that due to the arbitrary spreading of users in the area of MBS, the model is general, and the assumption of requesting one content by each user is not restrictive. 
 Denote $D=[d_{1}, ... , d_{Z}]$ as the demand vector of all users, where $d_{i}$ is the content requested by user $i$. 
 Based on all cached pieces of contents, requests of all users, and network topology,  the MBS sends the required files over the shared link to satisfy these requests. The traffic load in the delivery phase, denoted by $R$,  to satisfy all requests is a random variable. Unlike the placement phase, in the delivery phase, the cost of network load is high since the available bandwidth of the shared link is the system's bottleneck, so we only consider the traffic load in the delivery phase.
\begin{table}
	\centering
	\small
	\caption{Summery of the main notation.}
	\label{table:notation}
\begin{tabular}{|l|l|}
	\hline 
	Symbol & Explanation \\ 
	\hline 
	$n$ & generic content  \\ 
	$c$ & generic SBS (cache)  \\ 
	$i$& generic step of sending coded contents \\
	$p$ & generic cluster of SBSs\\
	$N$ & number of contents  \\ 
	$F$ & size of each content(bit)  \\
	$K$ & number of SBS \\ 
	$M$ & cache capacity of SBSs (content)  \\
	$Z$ & number of users (demands) in the range of MBS \\ 
	$q_{c,j}$ & probability of requesting the $j$th distinct\\&
	coded content at the next request in SBS $c$ \\
	$p_{n}$ & popularity of content $n$ \\
	$Q_i$ & number of non-empty coded queues  at step $i$\\
	$l^{(z)}_{c}$ &  number of distinct coded request in first $z$\\ &
	 requests that received by SBS $c$\\
	$Y_{n,c}$& indicate content $n$ is (or not) cached in SBS $c$ \\ 
	$X_{n,p}$& indicate content $n$ is (or not) cached in cluster $p$\\
	$S_{c,p}$& indicate SBS $c$ is (or not) member of cluster $p$ \\
	$r_1$& expected traffic load (MBS) for the coded contents\\ 
	$r_2$& expected traffic load (MBS) for un-cached contents\\ 
	$r$& total expected traffic load of the MBS $(r_1+r_2)$\\ 
	\hline 
\end{tabular} 
\end{table}

\section{Proposed Scheme}
\label{PS}
In this section, we first introduce our proposed $MAHC$ a multi-access hybrid coded-uncoded scheme, which is based on the baseline hybrid coded-uncoded caching scheme \cite{ghaffarisheshjavani2021content,9120820} for multi-access scenarios and formulate the optimal caching strategy. Then, we analyze the traffic load on the shared medium in the case of 2-SBS scenarios.
\subsection{The proposed Multi Access Hybrid coded-uncoded Caching ($MAHC$)}
As mentioned before, in the baseline hybrid coded-uncoded caching scheme, each cache is divided into two coded and uncoded parts. Then, the contents are sorted based on their popularity, and the most popular ones are selected to be cached on the uncoded part entirely. Some other popular contents also are selected to be cached based on the original coded caching scheme in the coded part of the caches. There is a trade-off between diversity and popularity in the hybrid coded-uncoded caching to select the best caching strategy (i.e., how to partition caches between coded and uncoded parts and determining the number of contents that should be selected to cache in the coded part). 
 In multi-access caching finding the optimal caching strategy is more challenging. In this case, the network topology can affect the trade-off between diversity and popularity. Therefore, we modified the baseline hybrid coded-uncoded scheme to be used in multi-access scenarios in $MAHC$ as follows:

In the placement phase of $MAHC$, same as the baseline hybrid caching, the cache memory of each SBS $c$ is divided into two parts: one part for uncoded and the other part for the coded caching. Unlike the baseline hybrid coded-uncoded caching scheme, in $MAHC$, different contents can be cached in the uncoded part of different caches. In addition, we have considered the possibility of the existence of several coded delivery clusters, where contents that are selected to be cached with the coded scheme can be different in different clusters. In the following, we will give a precise definition of the clusters and indicate their role in final optimization problem.

Let disjoint subsets $P = \{P_1,P_2,...,P_{|P|}\}$ define a partition of $S$ (the set of SBSs),  where $|P_i|\geq 1 $,  and ${\cup}^{|P|}_{i=1} P_{i} = S$. According to this definition, we have $|P|$ disjoint clusters of SBSs in partition $P$. A cluster must have at least two members to apply the coded caching scheme. Therefore, if there is only one SBS in a cluster, the entire cache memory of this SBS is allocated for uncoded caching. let $S_{c,p}=1$ if SBS $c$ is a member of cluster $p \in P$, and $S_{c,p}=0$ otherwise (where $\forall c \in \{1,2,\ldots,K\}$ we have $\sum_{p\in P}{S_{c,p}} = 1  $). Then, the number of SBSs in cluster $p$, denoted by $K_p$, is derived as $\sum_{c}  S_{c,p}=K_p$. A separate coded caching scheme is applied in each cluster $p$ in which $K_p \geqq 2$. SBSs which are the members of cluster $p$ dedicate the capacity $M_p \times F$ bits ($M_p \in [0, M]$) of their cache for the coded scheme. Moreover, the remaining $(M-M_{p}) \times F$ bits of their cache are allocated for the uncoded caching. 

In the delivery phase of $MAHC$, each user requests a content. Based on the location of the user, its requested content is in one of the following categories: 
\begin{itemize}
	\item $Uncoded\ cached  \ contents$: If the requested content is cached entirely (in the uncoded part) in one of the caches that the user has access to them, this request is served by this cache and thus, does not impose any traffic load on the shared medium.
	\item  $coded\ cached  \ contents$:  If the requested content is not cached entirely in any of the caches that the user has access to them, but it is partially cached based on the coded schemes in one or more caches, the cached parts are served to the user by the corresponding caches and other un-cached parts are served by the MBS using the coded scheme through the shared medium.
	\item  $ Un-cached \ contents$: If the requested content is not cached in any of the caches that the user has access to, then it should be served entirely by the MBS through the shared medium. Note that in this case all other coded requests for this content which are issued by other users who have partial access to parts of that content will be ignored. 
\end{itemize}
Therefore, the MBS is involved in delivering un-cached and coded cached contents. At first, it serves the requested contents from the category of un-cached contents on the shared medium. Then, it handles requests in the category of coded cached contents by applying the coded scheme within each coded delivery cluster separately. It is worth noting that the contents requested by at least one user in un-cached form will not be sent via the coded scheme.

In coded caching, diverse pieces of files are cached in different caches. In the baseline coded caching scheme, contents pieces are combined (XoRed) with each other and create a subfile. Therefore, although users do not need to receive the piece of content that they have access to, the corresponding subfile is not sent only when all pieces that create it are not needed by corresponding users. In other words, a subfile does not need to send on the shared medium only if there is no need for all pieces that created it. Therefore, in the baseline coded caching, accessing users to more pieces of files does not remarkably reduce the traffic load on the shared medium, especially when $K$ is large. In addition, the proposed $MACC$ schemes in previous works only consider specific topologies, which are not realistic in general. Also, we can apply the coded scheme of $MACC$ in the coded part of $MAHC$ given that the network topology is the same as the one considered in $MACC$.

\subsection{Problem Formulation}
 According to previous works, we assume that the placement phase is carried out during off-peak times, and therefore the cost of traffic load is negligible in this phase. In contrast, the available bandwidth is the bottleneck in the delivery phase and leads to a considerable cost. Hence, we ignore the traffic load in the placement phase and our goal is to minimize the traffic load over the shared link during the delivery phase. Let $Y_{n,c}=1$ if content $n$ is cached uncoded at SBS $c$ and  $Y_{n,c}=0$ otherwise. Also, let $X_{n,p}=1$ if content $n$ is selected to be cached coded in the memories of all SBSs in cluster $p$ and $X_{n,p}=0$, otherwise (i.e., if $X_{n,p}=1$ then $W_n$ is cached with the coded scheme in the cache of all SBSs $c$ where$S_{c,p}=1$). Consequently, the number of contents participating in the coded scheme of cluster $p$, denoted by $N_p$, is derived as $\sum_n X_{n,p}=N_p, \forall p$. Let $\widetilde{P}$ denote the set of all possible partitioning of SBSs $|\widetilde{P}|= B_K$ where $B_K=\sum_{i=0}^{K}{{K}\choose {i}}\times B_{i-1}$ is the total number of partitions of a set of size $K$. Let $r$ be the expected traffic load over the shared link in the delivery phase, i.e., $r=E[R]$. Also, consider $r_1$ and  $r_2$ as the expected traffic rates sent over the shared link at the delivery phase by the MBS to satisfy requests from the coded category and un-cached category from all clusters, respectively. Therefore, we have $r=r_1+r_2$, and our goal is to optimize the content placement in the SBSs' caches i.e., optimize clustering, cache partitioning, and content selection to cache in order to minimize the expected traffic load over the shared link during the delivery phase (i.e., $r$). Thus, the optimization problem is:
\begin{equation}
\begin{aligned}
\min_{\mathcal{P} \subset \mathcal{\widetilde{P}}} \quad & \{ \min \limits_{\substack{0 \leq M_{p} \leq  M,\vspace{0.1em}\\ M_p < N_p \leq N,
\vspace{0.1em}\\ S_{c, p} , X_{n, p} , Y_{n,p} \in \{0,1\}}}  r \} \\
\vspace{0.5em}
\textrm{s.t.} \quad & \sum_{n=1}^{N}\!{X_{n, p} } = N_p, \quad \forall p \in \mathcal{P},\\
& \sum_{n=1}^{N}\!{Y_{n,c} } = M-\sum_{p\in P} S_{c,p}\times M_{p} , \quad \forall c \in \{1,2,\ldots,K\},    \\
& \! \sum_{p\in P}{ S_{c,p}} = 1 , \quad \forall c \in \{1,2,\ldots,K\}.
\end{aligned}
\label{optimization}
\end{equation}
In general, the optimization problem of conventional and $MACC$ schemes are special cases of this optimization problem. 
\subsection{Performance Analysis}
\begin{figure}[!t]
	\centering
	\includegraphics[width=3.2in]{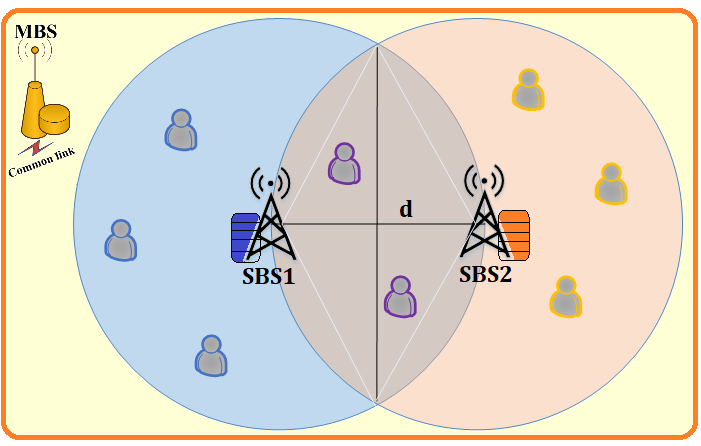}
	\caption{Network topology for two cache scenario with $Z$ users.}
	\label{TwoSBS}
	\vspace{-0.5em}
\end{figure}

 In the baseline hybrid caching, the best caching strategy, i.e., selecting the best contents to cache in each cache memory, depends on the system parameters such as contents popularity, the number of caches, memory capacity of caches, and so on. To select the best caching strategy in $MAHC$, the network topology should also be considered. In particular, a general formula can be written for the traffic load over the shared link for delivering un-cached requested contents (i.e., $r_2$), but calculating the traffic load for delivering the requested coded contents (i.e., $r_1$) highly depends on the exact network topology in multi-access scenarios. 
In addition, because of non-uniform content popularity distribution and multi-access scenario, finding the optimal placement strategy for the $MAHC$ and $MACC$ for large networks 
is not straightforward. Therefore, in this paper, we have chosen a simple but general network topology that includes two SBSs as shown in \figurename{~\ref{TwoSBS}}, and analyze the performance of the proposed $MAHC$ and other baseline $MACC$ schemes for this topology.

We assume that all $Z$ users are uniformly spread in the area of two SBSs. Therefore, if $Z_c$ is the number of users in the coverage area of SBS $c$, then $E[Z_c]=(S_c/S_{union})\times Z$, where $S_c$ is the area of SBS $c$ and $S_{union}$ is the area of the union of SBSs.
In this 2-SBSs network topology, users in the intersection area of two SBSs have access to the cache of both SBSs and other users have access to only one cache. The area of intersection depends on the radius of SBSs' areas and the distance between them(i.e., $d$ in \figurename{~\ref{TwoSBS}}). Therefore, all users access both caches if $d=0$ and the radius of two SBSs are equal. On the contrary, if $d$ is greater than the sum of two SBSs' radius, then the total area of intersection equals zero, and all users have only access to one cache. In this topology, partition $P=\{ \{ SBS1\}, \{ SBS2\}\}$ is a special case of partition $P^{\prime}=\{ \{ SBS1, SBS2\}\}$ when $M_p=0$. Therefore, only partition $\{ \{ SBS1, SBS2\}\}$ should be considered for this topology, which has only a cluster that includes two caches. 

In two caches $MACC$ scenarios, users who have access to both caches (i.e., users located in the intersection area of two SBSs) have access to all coded cached contents pieces. Therefore in this topology, if these users request for contents cached with the coded scheme, they can receive contents pieces from local SBSs' cache, and therefore delivering requested contents to them does not impose any load on the MBS. Therefore, In 2-SBS topologies, we call requests for coded-cached contents from users who are not located in the intersection area as $coded\ requests$. 

In the following, we derive the traffic load of the MBS under the proposed $MAHC$ for the above two SBSs network topology under non-uniform content popularity distribution.
\begin{thm}
\label{lemma1}
Let $P_{c,i}$ denote the probability that the number of distinct coded requests from the users who have access only to SBS $c$, 
denoted by $l_{c}$, is equal to or greater than $i$\footnote{$i$ refers to the step $i$ of coded caching scheme (please refer to Section \ref{hybrid_Back} for more detail) } (i.e., $P_{c,i}=Pr\{l_{c}\geq i\}$. $P_{c,i}$), is derived as follows:
\begin{align}
&P_{c,i}=  \sum_{j=i}^{Z}{Pr\{ l_{c}=j\}}.
\label{lem5-4}
\end{align}
Also, let $ \text{Pr}\{ l_{c}^{(z)} = j \} $ be the probability of having $j$ distinct coded requests in the first $z$ requests from the users who have access only to SBS $c$, 
 where $z \!=\! 1, 2,\ldots, Z$, then:
\begin{align}
&\text{Pr}\{ l_{c}=j \}= \text{Pr}\{ l_{c}^{(Z)} = j \},
\end{align}
where $\text{Pr}\{ l_{c}^{(Z)} = j \}$ can be calculated with below recursive formula:
\begin{align}
&1) {Pr}\{ l_{c}^{(0)} = 0 \}=1 ,  {Pr}\{ l_{c}^{(z)} = j | j>z \}=0,
\nonumber \\
& 2) {Pr}\{ l_{c}^{(z)} = 0 \} ={Pr}\{ l_{c}^{(z-1)} = 0 \} \times (1-q_{c, 1}),
\nonumber \\   
&3) {Pr}\{ l_{c}^{(z)}=j\} = {Pr}\{ l_{c}^{(z-1)}=j \} \times (1-q_{c, j+1})
\nonumber \\
& +{Pr}\{ l_{c}^{(z-1)}=j-1 \} \times q_{c, j},
\label{lem5-7} 
\end{align}
where $q_{c, j} $ is approximated to be:
\begin{align}
q_{c, j} \left\{ \begin{array}{l}
=0 \quad\quad\quad\quad\quad\quad\quad\quad \quad\quad\quad\quad\quad\quad \quad\quad if \quad j>N_p,\\ 
\simeq(1- \dfrac {j-1}{N_p} )\times\sum_{n=1}^{N}{X_{n,p}\times p_{n}}\times v_{c} \quad otherwise. 
\end{array}\right.
\label{lem5-8}
\end{align}
where 
\begin{align}
v_{c}= \dfrac{S_{c}-S_{Intersection}}{S_{union}}.
\label{lem5-v}
\end{align}
\end{thm}
\begin{proof}
	See Appendix A for the proof.
\end{proof}
\begin{prop}
\label{prop1}
	Let $Q_i$ be the random variable denoting the number of coded requests from all users in the coded delivery at step $i$, where $i=1, 2, \dots Z$. Note that $Q_i \in \{0,1,2\}$ since at each step only one coded request from each SBS can be responded. If $T=\frac{2 \times M_p}{N_p}$, then the expected traffic load over the shared link for delivering the coded requests, denoted by $r_1$, is:
	\begin{align}
	&r_{1} =  \begin{cases}
	 \scalebox{.95}{$ F\times \frac{2-T}{T+1}\times \sum_{i=1}^{Z} (1- Pr\{Q_{i}=0\})
	,\quad if\ N_p>M_p$}, \\
0,\quad \quad\quad\quad \quad\quad\quad\quad\quad\quad \quad\quad\quad\quad\quad   otherwise
		\end{cases}   
	\label{prop1-r1}
	\end{align}
where $Pr\{Q_{i}=0\}$ is the probability that there are no requests for coded contents at step $i$ from users who have only access to one of the two SBSs, and equals:
\begin{align}
&\text{Pr}\{ Q_{i}= 0 \}= \prod_{c=1}^2(1-P_{c,i}) ,
\label{prop1-q}
\end{align}
where $P_{c,i}$ is derived from lemma \ref{lemma1}.

 Moreover, let $S_{n}^{cached}$ be the total areas in which users have access to the caches that contain (coded or uncoded) content $n$. The expected traffic load over the shared link for delivering the un-cached content requests, denoted by $r_2$, is:\vspace{-0.4em}
	\begin{align}   
	r_{2}= F\times\sum_{n=1}^{N} \Bigg(1- \Big(1-(p_{n}\times (1-\dfrac{S_{n}^{cached}}{S_{union}}) )\Big)^{Z}\Bigg).
	\label{prop1-r2}
	\vspace{-0.4em}
	\end{align}
	where 
	\begin{align}
	S_{n}^{cached} = \left\{ \begin{array}{l}
	S_{union}  \quad\quad \quad \quad \quad if ( X_{n,p}=1 \ || \  \sum_{c} Y_{c,n}=2) ,\\ 
	\sum_{c} Y_{c,n}\times S_c  \quad \quad if (\sum_{c} Y_{c,n}=1),
	\\ 
	 0 \quad \quad\quad \quad\quad\quad \quad otherwise.  
	\end{array}\right.
	\label{prop1-S}
	\end{align}
Finally, the total expected traffic load is $r = r_1+r_2$.
\end{prop}
\begin{proof}
	Proof of equation \eqref{prop1-r1}: If $N_p \le M_p$ then all coded contents are cached in each cache completely, and therefore all users have access to all coded contents and can receive them without any traffic load over the shared link. In two cache scenarios, if $N_p > M_p$ then only users who have access to both caches receive their coded requests from caches without any traffic load over the shared link, and coded requests from users who access only one cache impose a traffic load on the shared link. According to the original coded caching scheme in \cite{maddah2014fundamental}, each content is split into ${{K}\choose {T}}$ non-overlapping fragments, each of size $F/{{K}\choose{T}}$. Each cache selects $T/K$ of all fragments and caches them. Based on the original coded caching scheme, $K$ coded requests (from users who have access to $K$ different caches) can be satisfied by sending ${K}\choose{T+1}$ subfiles, each of size $F/{{K}\choose{T}}$ over the shared link, where each subfile is created by $XOR$ of $T+1$ fragments of the requested contents. For two caches scenarios, two coded requests from users who do not have access to the same caches can be satisfied by sending $F \times{{2}\choose{T+1}} /{{2}\choose{T}} =  F \times \frac{2-T}{T+1}$ bits over the shared link. 
	
	If there are coded requests only from users who have access to the same cache, because it can not $XOR$ them, satisfying each one of these coded requests also imposes $F \times \frac{2-T}{T+1}$ bits traffic load over the shared link. Therefore, the expected traffic load over the shared link to satisfy coded requests (i.e., $r_1$) is proportional to the expected number of coded delivery steps where at each step at least one (one or two) coded request is satisfied. $Pr\{Q_i>0\}$ is the probability that the number of coded requests in the coded delivery step $i$ is greater than $0$, and it equals the probability that at least one of the two SBSs receives at least $i$ distinct coded requests from users who have only access to it. Therefore, if $H$ denotes the number of coded delivery steps, then $Pr\{H \ge i \}= Pr\{Q_{i}>0\} \nonumber$, and we have:
	 \begin{align}
	 	r_{1} &= E[F\times \frac{2-T}{T+1}\times H]	\nonumber \\
	 	&=  F\times \frac{2-T}{T+1}\times E[H]	\nonumber \\
 		&=  F\times \frac{2-T}{T+1}\times \sum_{i=1}^{Z} (Pr\{H \ge i \})\nonumber\\
	 	&=  F\times \frac{2-T}{T+1}\times \sum_{i=1}^{Z} (Pr\{Q_{i}>0\})\nonumber\\
	 	&=  F\times \frac{2-T}{T+1}\times \sum_{i=1}^{Z} (1- Pr\{Q_{i}=0\})		 		
	 	\label{prop1-H}
	 \end{align}
	where, the maximum number of steps equals $Z$ (i.e., $max(H)=Z$), which happens if all $Z$ users have access to the same cache and all of them request for coded contents.	 	
	
	Proof of equation \eqref{prop1-q}: $Pr\{Q_{i}=0\}$ is the probability that the number of coded requests in the coded delivery step $i$ equals $0$, and it equals the probability that both SBSs receive less than $i$ distinct requests for coded contents from users who only access one of them. Since the area of SBSs may be different, in lemma \ref{lemma1} we define $P_{c,i}$ as the probability that SBS $c$ receives equal or greater than $i$ distinct requests for coded contents from users who only have access to it. Therefore, the probability that both SBSs receive less than $i$ distinct requests for coded contents from users who only access one of them (i.e., $Pr\{Q_{i}=0\}$) equals $\prod_{c=1}^2(1-P_{c,i})$.
		
	Proof of equation\eqref{prop1-r2}: as mentioned before, $r_2$ is the traffic load for satisfying requests for un-cached (no coded and no uncoded) contents. But, The caching status of contents is different in different locations. For example, a particular content may be cached in $SBS_1$ but not in $SBS_2$ and vice versa. The probability that a user requests content $n$ is $p_n$ and the probability that a user has access to caches that cached content $n$ equals ${S_{n}^{cached}}/{S_{union}}$ 
	We assume that these two events (requesting content and users' locations) are independent. Therefore, the probability that a user requests content $n$ and is located in the coverage area that does not access caches which cache content $n$ equals $p_n \times (1-({S_{n}^{cached}}/{S_{union}}))$. Moreover, MBS should send an un-cached content over the shared link, if it is requested at least once (at least from one user in the total area of all SBSs). In other words, if a specific content is requested by multiple users in the delivery phase, due to the broadcast nature of the shared link, the MBS needs to send the content only once over the shared link. The probability that an event happens at least once can be found by subtracting the probability that the event does not happen from 1. Hence, the probability that from all users in the area of MBS, content $n$ is requested at least by one user who does not have access to it in SBSs' caches equals to 	
	$1-\Big(1-(p_n \times (1-\dfrac{S_{n}^{cached}}{S_{union}}))\Big)^{Z}$. 
	The expected traffic load to satisfy requests for all un-cached contents ($r_2$) is proportional to the expected number of distinct requests for them. The expected total number of distinct requests of un-cached contents is equal to the sum of the aforementioned expected probability for all un-cached contents, as indicated in \eqref{prop1-r2}. 
	
	Equation \eqref{prop1-S} indicates that $S_{n}^{cached} = S_{union}$ when content $W_n$ is cached with the coded scheme (i.e., $X_{n,p}=1$) or cached with the uncoded scheme in both caches (i.e., $\sum_{c}Y_{n,c}=2$). Otherwise, if content $W_n$ is cached with the uncoded scheme in only one SBS (i.e., $\sum_{c} Y_{c,n}=1$), then $S_{n}^{cached}$ equals the coverage area of that SBS. Finally, if content $W_n$ is not cached at all, (i.e., $X_{n,p}=0 \ and \ \sum_{c}Y_{n,c}=0$), then $S_{n}^{cached} =0$. This completes the proof.	
\end{proof}
We now formulate the optimum partitioning problem in order to minimize the traffic load over the shared link, i.e., $r$. 
The minimization problem is formulated as follows:
\begin{equation}
\label{optimizationTwoSBS}
\begin{aligned}
 & \min \limits_{\substack{0 \leq M_{p} \leq  M,\vspace{0.1em}\\ M_p < N_p \leq N,
		\vspace{0.1em}\\ X_{n, p} , Y_{n,p} \in \{0,1\}}}  r  \\
\vspace{0.5em}
\textrm{s.t.} \quad & \sum_{n=1}^{N}\!{X_{n, p} } = N_p, \\
& \sum_{n=1}^{N}\!{Y_{n,c} } = M- M_{p} , \quad \forall c \in \{1,2\}.    \\
\end{aligned}
\end{equation}

Finally, the optimal value of \eqref{optimizationTwoSBS} can be found by exhaustive search.
\section{Numerical Results }
\label{NR_Two}
\label{sec6}
In this section, the performance of the proposed $MAHC$ scheme is evaluated and compared with the baseline $MACC$ and conventional uncoded caching schemes for the two-cache scenarios through numerical evaluations and simulation. 

We assume that the coverage area of each SBSs is a circle and that both coverage areas have the same radius.
The content popularity distribution follows the Zipf popularity profile with parameter $\alpha > 0$ as follows:
\begin{equation}
p_{n}=\dfrac{(\frac{1}{n})^\alpha}{\sum_{j=1}^{N}(\frac{1}{j})^\alpha}.
\label{eq1}
\end{equation}
We validate analytical results with the simulations conducted using MATLAB. Simulations repeated $2000$ times, wherein each time, $Z$ users are spread in the area of two SBSs based on the 2D uniform distribution, and each user requests one content based on the content popularity distribution. Finally, the mean traffic load over the shared link and $95\% $confidence intervals (i.e., $Mean \pm 1.96 \times Std/ \sqrt{2000}$) of these $2000$ randomly generated tests are reported.
 
In \figurename{~\ref{fig3}},  we suppose that the number of available contents in the MBS is $N = 10$, and the cache capacity of both SBSs is $M=3$. The number of users in the system in \figurename{~\ref{fig3-a} and ~\ref{fig3-b}} is $Z=10$, and it is variable in \figurename{~\ref{fig3-c} from $4$ to $16$. The content popularity distribution parameter is $\alpha=1.2$ in \figurename{~\ref{fig3-b} and ~\ref{fig3-c}}, and it is variable in \figurename{~\ref{fig3-a} from $0.7$ to $2$.
The ratio of the intersection area of SBSs (i.e., $S_{intersection}$) to the total area of SBSs (i.e., $S_{union}$) is variable in \figurename{~\ref{fig3-b} from $0$ to $1$, and it is equal to $0.3375$ (about $1/3$) in \figurename{~\ref{fig3-a} and ~\ref{fig3-c}} (i.e., expectedly about $1/3$ of users access to both SBSs and other $2/3$ of users access only one SBS).
		
\figurename{~\ref{fig3-a}} illustrates the simulation and analytical results for the traffic load over the shared link as a function of the Zipf parameter in the interval $\alpha \in [0.7, 2]$. As can be seen in this figure, the simulation results are very close to the analytical findings, and the proposed $MAHC$ scheme significantly offloads more traffic compared to the $MACC$ and the conventional uncoded schemes. In addition, it can be seen that for this network configuration, the $MACC$ scheme offloads more traffic than the conventional uncoded scheme when $\alpha$ is lower than $1.2$, and the performance (and optimal caching policy) of it is equal to our proposed $MAHC$ when $\alpha$ is lower than $0.8$.

\figurename{~\ref{fig3-b}} also depicts the traffic load over the shared link as a function of the ratio of $S_{intersection}/S_{union}$. When this ratio becomes larger, it means that the intersection area becomes larger, and therefore more users have access to both SBSs. This ratio is directly related to the expected number of caches each user has access to and is used to study the impact of cache density per unit area on the caching performance. To this aim, we set a fixed value for the coverage radius of SBSs, and change the distance between them. For example, when the distance between two SBSs equals $0.8$ of the SBS's coverage radius, $S_{intersection}/S_{union}=0.3375$. As can be seen in this figure, when all users have access to both SBSs (i.e., $S_{intersection}/S_{union}=1$), the performance of all caching schemes is the same (for this configuration), but when a few users access only one SBS, $MAHC$ and $MACC$ have better performance than conventional uncoded caching. In addition, It is evident from this figure that the performance of the proposed $MAHC$ is significantly better than the two other schemes. 

\figurename{~\ref{fig3-c}} illustrates the traffic load over the shared link as a function of the number of users in the system ($Z$). This figure also shows that the proposed $MAHC$ outperformance the previous schemes for different numbers of users in the system. 
\begin{figure*} 
    \centering
  \subfloat{\includegraphics[width=0.32\linewidth]{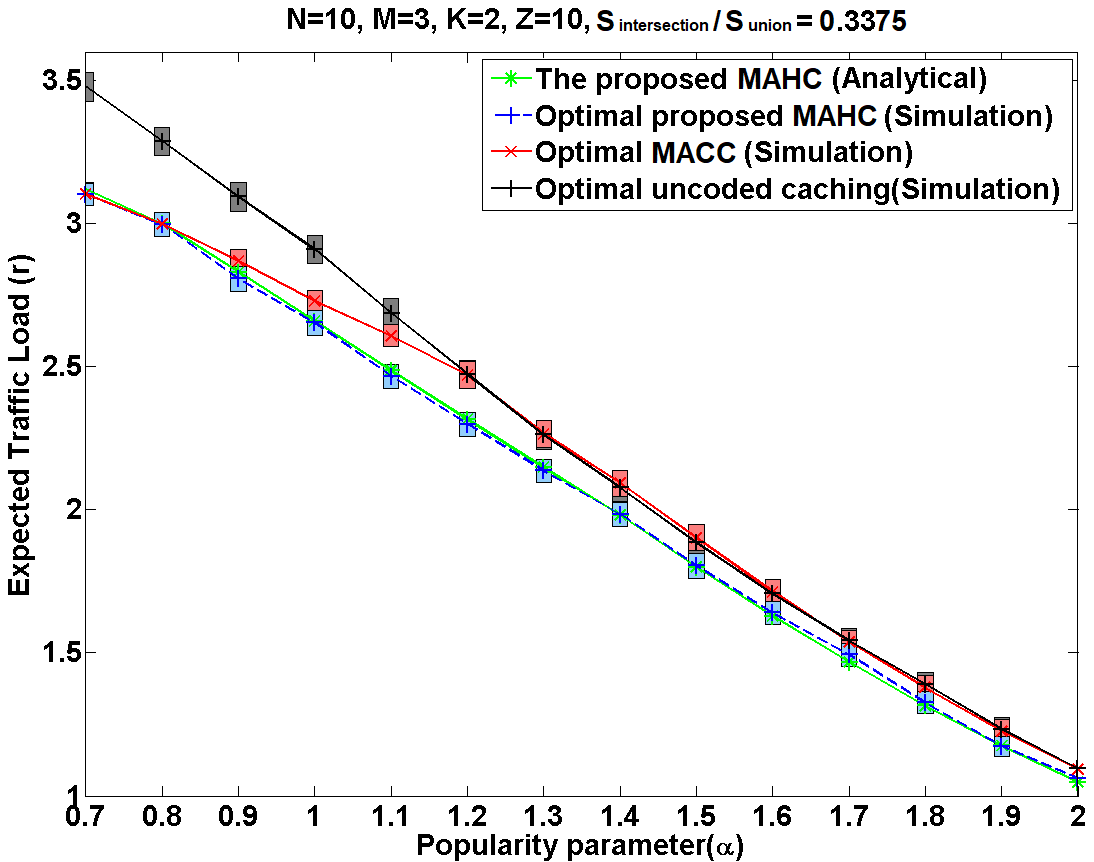} \label{fig3-a}}
     \,
  \subfloat{\includegraphics[width=0.32\linewidth]{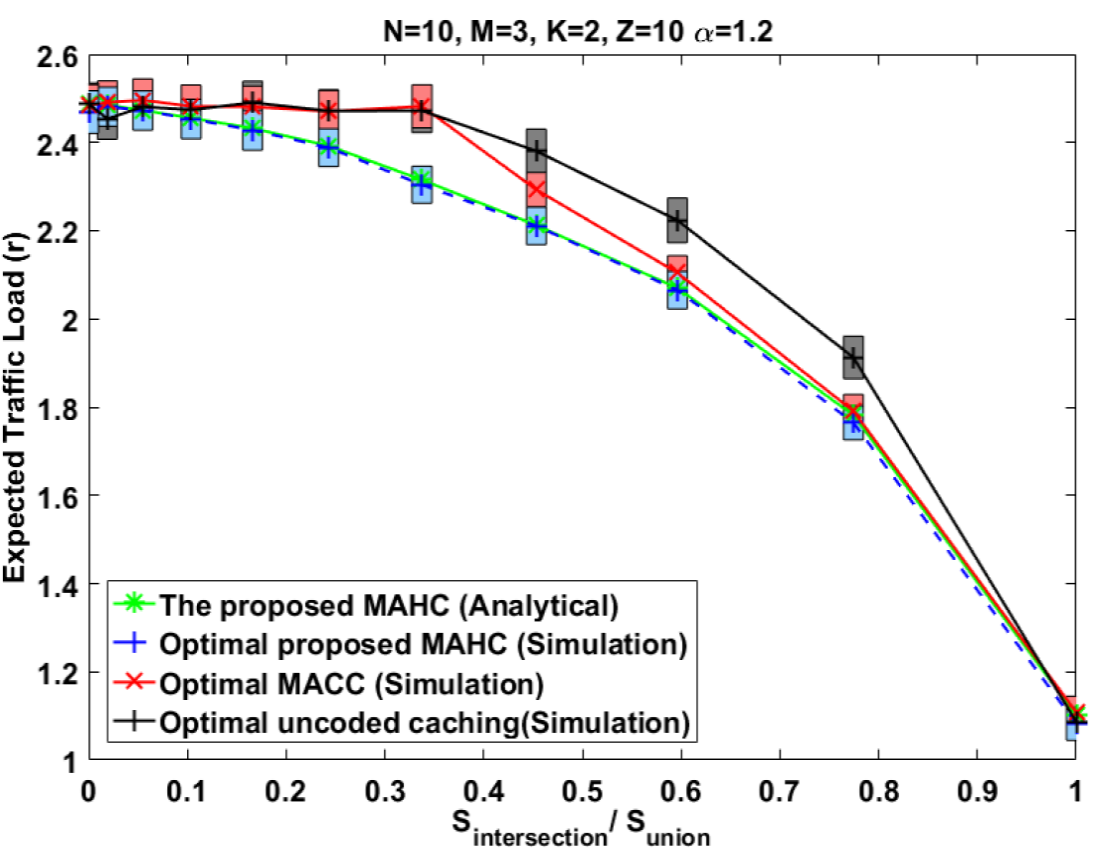}
\label{fig3-b}}
     \,
  \subfloat{\includegraphics[width=0.32\linewidth]{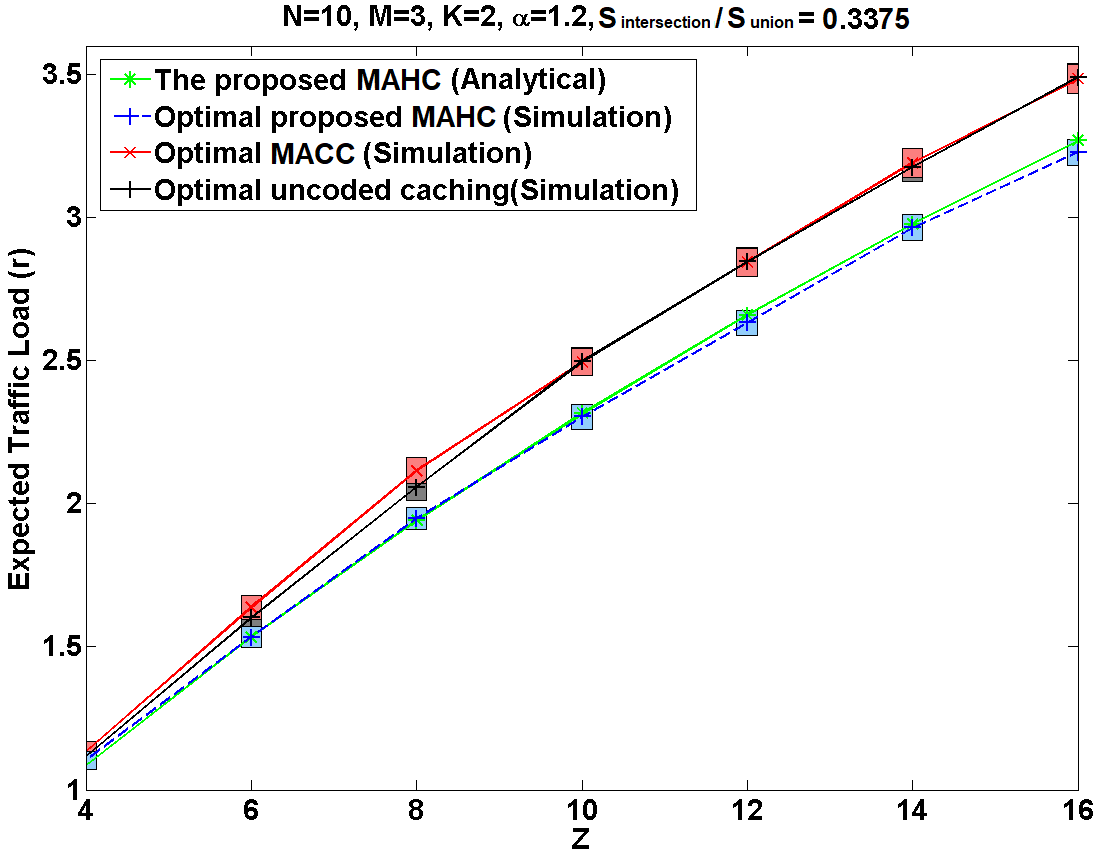}
\label{fig3-c}}\vspace{-0.5em}
  \caption{Expected traffic load over the shared link in the delivery phase versus (a)~the popularity parameter ($\alpha$), (b), the ratio of $S_{intersection}/S_{union}$ and (c)~the number of users in the system ($Z$), (results have been normalized to $F$, rectangles indicate the 95\% confidence intervals).}
  \label{fig3}
  \vspace{-0.5em} 
\end{figure*}
\section{Conclusion and future works}
\label{cocln}
In this paper, we have studied content caching under non-uniform content popularity distribution in general multi-access networks, where each user can access multiple cache nodes, and proposed the multi-access hybrid coded-uncoded caching ($MAHC$). 
In practice, the proposed scenario corresponds to a multi-access cellular network that includes an MBS which access to all contents library, and multiple users and SBSs were randomly spread in the area of the MBS's coverage and have access to the MBS through a shared link. Each SBS is equipped with a limited-size cache and each user could access multiple SBSs.  
Our goal is to minimize the traffic load over the shared link, which is the system's bottleneck. 
We formulate the problem for the proposed $MAHC$ in general multi-access network topology. However, calculating the exact traffic load over the shared link for coded delivery highly depends on the exact network topology. 
Therefore, in this paper, we consider a simple but general network topology that includes two SBSs. Validated by simulation results, our findings showed that the proposed $MAHC$ scheme significantly outperforms the $MACC$ and the uncoded schemes under non-uniform content popularity distribution.
For future topics, we plan to propose heuristic methods in the case of more complicated network topologies to find appropriate caching strategies in polynomial complexity.

\appendices
\section{Lemma1Proof}
\begin{proof}
	If $l_{c}$ denotes the number of distinct coded requests from users who have only access to the SBS $c$, then $P_{c,i}$ equals $Pr\{l_{c}>=i\}$ and can be calculated according to \eqref{lem5-4}. 
	Due to the possibility of requesting the same coded content several times by different users, the native approach for calculating $Pr\{l_{c}=j\}$ has exponential complexity. However, based on the area of SBSs, an approximation of it can be calculated with recursion and dynamic programming in polynomial complexity as follows:
	If the $Pr\{l^{(z)}_{c}=j\} $ denotes the probability of having $j$ distinct coded requests from users who have only access to the SBS $c$ in first $z$ requests, where $z \!=\! 1, 2,\ldots, Z$, then $Pr\{l_{c}=j\}$ is equal to  $Pr\{l^{(Z)}_{c}=j\}$ and can be calculated with a recursive formula that is given in \eqref{lem5-7}. In this equation, the notation $q_{c,j}$ denotes the probability of requesting the $j$th distinct coded content from users who have only access to the SBS $c$ at the next request, where until this request $j-1$ distinct coded contents have been requested by users who have only access to the SBS $c$. Therefore, $q_{c,j}$ is independent of the total number of requests, and it depends on the popularity of coded contents and previous $j-1$ distinct requested coded contents, and the probability that the requested content is from the users who only have access to SBS $c$.
		
	Due to the non-uniform contents popularity distribution and the possibility of requesting one of the $j-1$ previously requested coded contents by remaining users, calculating the exact amount of $q_{c,j}$ is highly complicated. To this aim, $q_{c,j}$ is approximated by the probability of requesting one of the $j-1$ previous requested contents as shown in \eqref{lem5-8}. In this equation, the array $X$ is used to determine coded contents, and $v_{c}$ denotes the probability that requested content is from the users who only have access to SBS $c$. We assume that users spread in the area of two SBSs uniformly, therefore, $v_{c}$ appropriate to the coverage areas of SBS $c$ which are not covered by the other SBS, and can be calculated by equation \ref{lem5-v}. We assume that the content popularity distribution($\{p_n\}_{n=1}^N$) and location of the users in the areas of both SBSs ($\{v_{c}\}_{c=1,2}$) are independent. Therefore, the joint probability of them (which equals $q_{c,j}$) is equal to the product of them as shown in \ref{lem5-8}.
	
	The approximation that is used in \eqref{lem5-8} is more accurate when the popularity distribution is almost uniform. In addition, when the popularity distribution is extremely non-uniform, where few contents are in high-demand, $MAHC$ tends to cache these high demand contents entirely (tends to a higher value for $M_P$). By caching the most popular contents entirely, the error of the approximation is greatly reduced. 
	Finally, it should be noted that we calculate $r$ for finding the best content placement rather than calculating its exact value. Hence, this approximation has a negligible impact on finding the best library partitioning, and content selections.
	This completes the proof.
\end{proof}
\bibliographystyle{IEEEtran}
\bibliography{journal} 

\begin{thebibliography}{10}
\providecommand{\url}[1]{#1}
\csname url@samestyle\endcsname
\providecommand{\newblock}{\relax}
\providecommand{\bibinfo}[2]{#2}
\providecommand{\BIBentrySTDinterwordspacing}{\spaceskip=0pt\relax}
\providecommand{\BIBentryALTinterwordstretchfactor}{4}
\providecommand{\BIBentryALTinterwordspacing}{\spaceskip=\fontdimen2\font plus
\BIBentryALTinterwordstretchfactor\fontdimen3\font minus
  \fontdimen4\font\relax}
\providecommand{\BIBforeignlanguage}[2]{{%
\expandafter\ifx\csname l@#1\endcsname\relax
\typeout{** WARNING: IEEEtran.bst: No hyphenation pattern has been}%
\typeout{** loaded for the language `#1'. Using the pattern for}%
\typeout{** the default language instead.}%
\else
\language=\csname l@#1\endcsname
\fi
#2}}
\providecommand{\BIBdecl}{\relax}
\BIBdecl

\bibitem{2022ericsson}
J.~Peter \emph{et~al.}, ``Ericsson mobility report,'' pp. 1--40, 2022.

\bibitem{chandrasekhar2008femtocell}
V.~{Chandrasekhar}, J.~G. {Andrews}, and A.~{Gatherer}, ``Femtocell networks: a
  survey,'' \emph{IEEE Communications Magazine}, vol.~46, no.~9, pp. 59--67,
  Sep. 2008.

\bibitem{shanmugam2013femtocaching}
K.~{Shanmugam}, N.~{Golrezaei}, A.~G. {Dimakis}, A.~F. {Molisch}, and
  G.~{Caire}, ``Femtocaching: Wireless content delivery through distributed
  caching helpers,'' \emph{IEEE Transactions on Information Theory}, vol.~59,
  no.~12, pp. 8402--8413, Dec 2013.

\bibitem{chen2017probabilistic}
Y.~{Chen}, M.~{Ding}, J.~{Li}, Z.~{Lin}, G.~{Mao}, and L.~{Hanzo},
  ``Probabilistic small-cell caching: Performance analysis and optimization,''
  \emph{IEEE Transactions on Vehicular Technology}, vol.~66, no.~5, pp.
  4341--4354, May 2017.

\bibitem{maddah2014fundamental}
M.~A. {Maddah-Ali} and U.~{Niesen}, ``Fundamental limits of caching,''
  \emph{IEEE Transactions on Information Theory}, vol.~60, no.~5, pp.
  2856--2867, May 2014.

\bibitem{9120820}
A.~Ghaffari~Sheshjavani, A.~Khonsari, S.~P. Shariatpanahi, M.~Moradian, and
  A.~Dadlani, ``Coded caching under non-uniform content popularity
  distributions with multiple requests,'' in \emph{2020 IEEE Wireless
  Communications and Networking Conference (WCNC)}, 2020, pp. 1--6.

\bibitem{ghaffarisheshjavani2021content}
A.~Ghaffari~Sheshjavani, A.~Khonsari, S.~P. Shariatpanahi, and M.~Moradian,
  ``Content caching for shared medium networks under heterogeneous users’
  behaviors,'' \emph{Computer Networks}, vol. 199, p. 108454, 2021.

\bibitem{9839153}
A.~Ghaffari~Sheshjavani, A.~Khonsari, S.~P. Shariatpanahi, M.~Moradian, and
  A.~Dadlani, ``Content caching in shared medium networks with non-uniform and
  user-dependent demands,'' in \emph{ICC 2022 - IEEE International Conference
  on Communications}, 2022, pp. 2501--2506.

\bibitem{8374848}
E.~Leonardi and G.~Neglia, ``Implicit coordination of caches in small cell
  networks under unknown popularity profiles,'' \emph{IEEE Journal on Selected
  Areas in Communications}, vol.~36, no.~6, pp. 1276--1285, 2018.

\bibitem{9638333}
F.~Rezaei, B.~H. Khalaj, M.~Xiao, and M.~Skoglund, ``Performance analysis of
  heterogeneous cellular caching networks with overlapping small cells,''
  \emph{IEEE Transactions on Vehicular Technology}, vol.~71, no.~2, pp.
  1941--1951, 2022.

\bibitem{multilevel2017}
J.~{Hachem}, N.~{Karamchandani}, and S.~N. {Diggavi}, ``Coded caching for
  multi-level popularity and access,'' \emph{IEEE Transactions on Information
  Theory}, vol.~63, no.~5, pp. 3108--3141, May 2017.

\bibitem{9036921}
K.~S. Reddy and N.~Karamchandani, ``Rate-memory trade-off for multi-access
  coded caching with uncoded placement,'' \emph{IEEE Transactions on
  Communications}, vol.~68, no.~6, pp. 3261--3274, 2020.

\bibitem{9774404}
K.~K.~K. Namboodiri and B.~S. Rajan, ``Improved lower bounds for multi-access
  coded caching,'' \emph{IEEE Transactions on Communications}, vol.~70, no.~7,
  pp. 4454--4468, 2022.

\bibitem{9511442}
M.~Cheng, K.~Wan, D.~Liang, M.~Zhang, and G.~Caire, ``A novel transformation
  approach of shared-link coded caching schemes for multiaccess networks,''
  \emph{IEEE Transactions on Communications}, vol.~69, no.~11, pp. 7376--7389,
  2021.

\bibitem{8989128}
B.~Serbetci, E.~Parrinello, and P.~Elia, ``Multi-access coded caching: gains
  beyond cache-redundancy,'' in \emph{2019 IEEE Information Theory Workshop
  (ITW)}, 2019, pp. 1--5.

\bibitem{9551929}
S.~Sasi and B.~S. Rajan, ``Multi-access coded caching scheme with linear
  sub-packetization using pdas,'' \emph{IEEE Transactions on Communications},
  vol.~69, no.~12, pp. 7974--7985, 2021.

\bibitem{10068305}
J.~Wang, M.~Cheng, Y.~Wu, and X.~Li, ``Multi-access coded caching with optimal
  rate and linear subpacketization under pda and consecutive cyclic
  placement,'' \emph{IEEE Transactions on Communications}, pp. 1--1, 2023.

\bibitem{9771663}
K.~K. Krishnan~Namboodiri and B.~Sundar~Rajan, ``Multi-access coded caching
  with demand privacy,'' in \emph{2022 IEEE Wireless Communications and
  Networking Conference (WCNC)}, 2022, pp. 2280--2285.

\bibitem{9834460}
K.~Wan, M.~Cheng, D.~Liang, and G.~Caire, ``Multiaccess coded caching with
  private demands,'' in \emph{2022 IEEE International Symposium on Information
  Theory (ISIT)}, 2022, pp. 1390--1395.

\bibitem{karamchandani2016hierarchical}
N.~{Karamchandani}, U.~{Niesen}, M.~A. {Maddah-Ali}, and S.~{Diggavi},
  ``Hierarchical coded caching,'' in \emph{2014 IEEE International Symposium on
  Information Theory}, June 2014, pp. 2142--2146.

\bibitem{shariatpanahi2016multi}
S.~P. {Shariatpanahi}, S.~A. {Motahari}, and B.~H. {Khalaj}, ``Multi-server
  coded caching,'' \emph{IEEE Transactions on Information Theory}, vol.~62,
  no.~12, pp. 7253--7271, Dec 2016.

\bibitem{7055939}
R.~{Pedarsani}, M.~A. {Maddah-Ali}, and U.~{Niesen}, ``Online coded caching,''
  \emph{IEEE/ACM Transactions on Networking}, vol.~24, no.~2, pp. 836--845,
  April 2016.

\bibitem{7999228}
M.~{Mohammadi Amiri}, Q.~{Yang}, and D.~{Gündüz}, ``Decentralized caching and
  coded delivery with distinct cache capacities,'' \emph{IEEE Transactions on
  Communications}, vol.~65, no.~11, pp. 4657--4669, Nov 2017.

\bibitem{7342961}
M.~{Ji}, G.~{Caire}, and A.~F. {Molisch}, ``Fundamental limits of caching in
  wireless d2d networks,'' \emph{IEEE Transactions on Information Theory},
  vol.~62, no.~2, pp. 849--869, Feb 2016.

\bibitem{8374865}
Y.~{Lu}, W.~{Chen}, and H.~V. {Poor}, ``Coded joint pushing and caching with
  asynchronous user requests,'' \emph{IEEE Journal on Selected Areas in
  Communications}, vol.~36, no.~8, pp. 1843--1856, Aug 2018.

\bibitem{8594642}
H.~{Xu}, C.~{Gong}, and X.~{Wang}, ``Efficient file delivery for coded
  prefetching in shared cache networks with multiple requests per user,''
  \emph{IEEE Transactions on Communications}, vol.~67, no.~4, pp. 2849--2865,
  April 2019.

\bibitem{niesen2017coded}
U.~{Niesen} and M.~A. {Maddah-Ali}, ``Coded caching with nonuniform demands,''
  \emph{IEEE Transactions on Information Theory}, vol.~63, no.~2, pp.
  1146--1158, Feb 2017.

\bibitem{li2017traffic}
T.~{Li}, M.~{Ashraphijuo}, X.~{Wang}, and P.~{Fan}, ``Traffic off-loading with
  energy-harvesting small cells and coded content caching,'' \emph{IEEE
  Transactions on Communications}, vol.~65, no.~2, pp. 906--917, Feb 2017.

\bibitem{ji2017order}
M.~{Ji}, A.~M. {Tulino}, J.~{Llorca}, and G.~{Caire}, ``Order-optimal rate of
  caching and coded multicasting with random demands,'' \emph{IEEE Transactions
  on Information Theory}, vol.~63, no.~6, pp. 3923--3949, June 2017.

\bibitem{zhang2018coded}
J.~{Zhang}, X.~{Lin}, and X.~{Wang}, ``Coded caching under arbitrary popularity
  distributions,'' \emph{IEEE Transactions on Information Theory}, vol.~64,
  no.~1, pp. 349--366, Jan 2018.

\bibitem{8863425}
S.~A. {Saberali}, L.~{Lampe}, and I.~F. {Blake}, ``Full characterization of
  optimal uncoded placement for the structured clique cover delivery of
  nonuniform demands,'' \emph{IEEE Transactions on Information Theory},
  vol.~66, no.~1, pp. 633--648, Jan 2020.

\bibitem{9834838}
M.~Zhang, K.~Wan, M.~Cheng, and G.~Caire, ``Coded caching for two-dimensional
  multi-access networks,'' in \emph{2022 IEEE International Symposium on
  Information Theory (ISIT)}, 2022, pp. 1707--1712.

\bibitem{katyal2021multi}
D.~Katyal, P.~N. Muralidhar, and B.~S. Rajan, ``Multi-access coded caching
  schemes from cross resolvable designs,'' \emph{IEEE Transactions on
  Communications}, vol.~69, no.~5, pp. 2997--3010, 2021.

\bibitem{9611394}
P.~N. Muralidhar, D.~Katyal, and B.~S. Rajan, ``Maddah-ali-niesen scheme for
  multi-access coded caching,'' in \emph{2021 IEEE Information Theory Workshop
  (ITW)}, 2021, pp. 1--6.

\bibitem{9839411}
F.~Brunero and P.~Elia, ``Fundamental limits of combinatorial multi-access
  caching,'' \emph{IEEE Transactions on Information Theory}, vol.~69, no.~2,
  pp. 1037--1056, 2023.

\end{thebibliography}


\begin{thebibliography}{1}
\providecommand{\url}[1]{#1}
\csname url@samestyle\endcsname
\providecommand{\newblock}{\relax}
\providecommand{\bibinfo}[2]{#2}
\providecommand{\BIBentrySTDinterwordspacing}{\spaceskip=0pt\relax}
\providecommand{\BIBentryALTinterwordstretchfactor}{4}
\providecommand{\BIBentryALTinterwordspacing}{\spaceskip=\fontdimen2\font plus
\BIBentryALTinterwordstretchfactor\fontdimen3\font minus
  \fontdimen4\font\relax}
\providecommand{\BIBforeignlanguage}[2]{{%
\expandafter\ifx\csname l@#1\endcsname\relax
\typeout{** WARNING: IEEEtran.bst: No hyphenation pattern has been}%
\typeout{** loaded for the language `#1'. Using the pattern for}%
\typeout{** the default language instead.}%
\else
\language=\csname l@#1\endcsname
\fi
#2}}
\providecommand{\BIBdecl}{\relax}
\BIBdecl

\bibitem{katyal2021multi}
D.~Katyal, P.~N. Muralidhar, and B.~S. Rajan, ``Multi-access coded caching
  schemes from cross resolvable designs,'' \emph{IEEE Transactions on
  Communications}, vol.~69, no.~5, pp. 2997--3010, 2021.

\end{thebibliography}
\end{document}